# Improve Myocardial Strain Estimation based on Deformable Groupwise Registration with a Locally Low-Rank Dissimilarity Metric

Haiyang Chen [a], Juan Gao [a], Zhuo Chen [a], Chenhao Gao [a], Sirui Huo [a], Meng Jiang [b, *], Jun Pu [b, *], and Chenxi Hu [a, *]

[a] National Engineering Research Center of Advanced Magnetic Resonance Technologies for Diagnosis and Therapy, School of Biomedical Engineering, Shanghai Jiao Tong University, Shanghai, China

[b] Division of Cardiology, Renji Hospital, School of Medicine, Shanghai Jiao Tong University, Shanghai, China

*Corresponding Authors:

Chenxi Hu (chenxi.hu@sjtu.edu.cn), 415 S Med-X Center, 1954 Huashan Road, Shanghai 200030, China

Jun Pu (pujun310@hotmail.com), Key Laboratory of Coronary Heart Disease, Division of Cardiology, Renji Hospital, School of Medicine, Shanghai Jiao Tong University, Shanghai 200127, China

Meng Jiang (jiangmeng0919@163.com), Key Laboratory of Coronary Heart Disease, Division of Cardiology, Renji Hospital, School of Medicine, Shanghai Jiao Tong University, Shanghai 200127, China

# Abstract

**Background:** Current mainstream cardiovascular magnetic resonance-feature tracking (CMR-FT) methods, including optical flow and pairwise registration, often suffer from the drift effect caused by accumulative tracking errors. Here, we developed a CMR-FT method based on deformable groupwise registration with a locally low-rank (LLR) dissimilarity metric to improve myocardial tracking and strain estimation accuracy.

**Methods:** The proposed method, Groupwise-LLR, performs feature tracking by iteratively updating the entire displacement field across all cardiac phases to minimize the sum of the patchwise signal ranks of the deformed movie. The method was compared with alternative CMR-FT methods including the Farneback optical flow, a sequentially pairwise registration method, and a global low rankness-based groupwise registration method via a simulated dataset (n = 20), a public cine data set (n = 100), and an in-house tagging-MRI patient dataset (n = 16). The proposed method was also compared with two general groupwise registration methods, nD+t B-Splines and pTVreg, in simulations and in vivo tracking.

**Results:** On the simulated dataset, Groupwise-LLR achieved the lowest point tracking errors (almost all $p < 0.05$) and voxelwise/global strain errors (almost all $p < 0.05$). On the public dataset, Groupwise-LLR achieved the lowest contour tracking errors (all $p < 0.05$), reduced the drift effect in late-diastole, and preserved similar inter-observer reproducibility as the alternative methods. On the patient dataset, Groupwise-LLR correlated better with tagging-MRI for radial strains than the other CMR-FT methods in multiple myocardial segments and levels.

**Conclusions:** The proposed Groupwise-LLR reduces the drift effect and provides more accurate myocardial tracking and strain estimation than the alternative methods. The method may thus facilitate a more accurate estimation of myocardial strains for clinical assessments of cardiac function.

# Background

Myocardial strain is a sensitive measure of the left ventricular mechanical function, which can be useful to aid the diagnosis and prognosis of various diseases [1–3]. Cardiovascular magnetic resonance-feature tracking (CMR-FT) estimates myocardial strains from routinely performed cine MRI images using feature tracking algorithms [4]. Compared with the reference standard tagging-MRI, CMR-FT does not need additional time-consuming sequences and facilitates retrospective strain analysis [5]. However, it is well known that CMR-FT based on regular cine does not estimate regional strain accurately [6, 7]. Since regional strain is often hypothesized to be more sensitive to the early progresses of cardiac diseases, a more accurate estimation of the regional strain with advanced feature tracking algorithms is highly desirable.

Current commercial software tools for CMR-FT often use optical flow or pairwise registration for tracking of the myocardium across adjacent cardiac phases [8]. However, studies have shown that these frame-by-frame tracking methods often cause an accumulation of tracking errors through the early frames. These errors become greater for later frames, which eventually cause inaccurate feature tracking and strain estimation, often referred to as the drift effect [9]. Several methods have been proposed to address this challenge. One method is based on drift compensation, which approximates the drift by a linear function and subtracts the trend from the estimated strain [10]. However, since the linearity of the drift is assumed only for simplicity, the compensation is highly empirical and does not truly eliminate the accumulative tracking errors. Another type of pairwise CMR-FT methods seeks to directly align every frame to the end-diastolic frame [9, 11]. Although such methods can avoid the drift effect, large deformations between distant frames such as end-diastole and end-systole can cause even more severe tracking errors [12]. In addition, there are also attempts to mitigate the errors by combining forward and backward registration passes [13–15]. However, the tracking errors caused by pairwise alignment still exist in both passes and are not guaranteed to be offset when combined. Recently, groupwise registration has shown the potential to reduce the drift effect by simultaneously aligning all cine images to a common reference frame. For example, Metz et al. proposed a deformable groupwise registration method for tissue tracking, which used a sum of the voxelwise signal variances as the dissimilarity metric since a well-aligned set of images should have a low signal variation across time [16]. Qiao et al. implemented a similar groupwise registration method for myocardial tracking via cine MRI images and demonstrated an improved tracking accuracy over pairwise registration [17]. In addition to signal variance, *signal low-rankness* across the spatiotemporal domain can also be used as a dissimilarity metric for groupwise registration. If the registration is perfect and the temporal signal variation is spatially consistent, the aligned images should encompass a low-rank property, which can be used to guide the registration. For example, Peng et al. decomposed the spatiotemporal images into an error term, which should be sparse, and a set of well-aligned images, which should have the low-rank property [18]. They then combined the sparsity and low-rankness into a single dissimilarity metric to guide the groupwise registration, and found improved accuracy and efficiency when compared with other registration methods. Haase et al. extended this method by introducing nonparametric deformation and total variation regularization, and found improved accuracy compared with alternative groupwise registration methods for several motion tracking applications [19]. However, none of these studies on groupwise registration evaluated the accuracy and reproducibility for strain estimation.

In this work, we sought to develop a strain estimation method based on deformable groupwise registration with a locally low-rank (LLR) dissimilarity metric and evaluate its accuracy and reproducibility. Previous myocardial tracking studies have shown that global low-rankness can be used to guide groupwise registration [18, 19]; however, it is well-known that signals in cine MRI are often locally low-rank rather than globally low-rank [20]. Different tissues in cine MRI may encompass signal variations of different patterns, and thus the global rank can be relatively high even after motion compensation. For example, the myocardial intensity may present a periodic variation across the cardiac cycle due to through-plane motion and the fresh spin effect. The blood signals may also change irregularly due to turbulent flows. The banding and flow artifacts commonly observed in cine MRI can disturb regional signals in the image. Therefore, Locally Low Rankness (LLR) may be a better metric to measure the alignment of cine images over the cardiac cycle. The baseline methods for evaluation included three alternative CMR-FT methods (optical flow, pairwise registration, and global low-rankness (GLR)-based groupwise registration) and two general groupwise registration methods (nD+t B-Splines [16] and pTVreg [21]). We compared our methods with all the baseline methods in terms of myocardial tracking accuracy via a simulated dataset (XCAT) [22] and a public cine dataset (ACDC) [23]. We further compared our method with the alternative CMR-FT methods in terms of strain estimation accuracy and reproducibility via the simulated dataset, public cine dataset, and an in-house tagging-MRI patient dataset.

## Methods

**Groupwise registration formulation**

The goal of groupwise registration is to simultaneously align all cine images to a common reference frame. Let $f(\mathbf{x}, t)$ represent the cine image sequence, where the spatial coordinate $\mathbf{x} \in \Omega = \{(x,y) | 1 \leq x \leq N_x, 1 \leq y \leq N_y\}$ and the temporal coordinate $1 \leq t \leq N_t$. $N_x$, $N_y$, and $N_t$ represent the pixel number in the x direction, pixel number in the y direction, and frame number, respectively. The deformed cine image sequence can be expressed as

$$\tilde{f}(\mathbf{x}, t) = f(\mathbf{T}(\mathbf{x}, t), t) = f(\mathbf{x} + \mathbf{d}(\mathbf{x}, t), t), \tag{1}$$

where $\mathbf{T}(\mathbf{x}, t) = \mathbf{x} + \mathbf{d}(\mathbf{x}, t)$ is the transformation field and $\mathbf{d}(\mathbf{x}, t)$ is the displacement field. We use the free-form deformation modeling of the displacement field [24]. Specifically, given a control point mesh $\boldsymbol{\phi}(i, j, t)$ with uniform control point spacing $\delta$, $\mathbf{d}(\mathbf{x}, t)$ is modeled by a linear combination of control points with cubic B-spline coefficients as

$$\mathbf{d}(\mathbf{x}, t) = \sum_{k=0}^{3} \sum_{l=0}^{3} B_k(u) B_l(v) \boldsymbol{\phi}(i+k, j+l, t), \tag{2}$$

where $i = \lfloor (x-1)/\delta \rfloor$, $j = \lfloor (y-1)/\delta \rfloor$, $u = (x-1)/\delta - \lfloor (x-1)/\delta \rfloor$, $v = (y-1)/\delta - \lfloor (y-1)/\delta \rfloor$, and $B_k(u)$ is the k-th basis function of the B-splines. By constraining the temporal average of the transformation or displacement field to be an identity map or zero, we register all images to an average common reference frame:

$$\frac{1}{N_t} \sum_{t=1}^{N_t} \mathbf{T}(\mathbf{x}, t) = \mathbf{x} \Leftrightarrow \frac{1}{N_t} \sum_{t=1}^{N_t} \mathbf{d}(\mathbf{x}, t) = \mathbf{0}, \tag{3}$$

which based on Equation (2) is equivalent to

$$\frac{1}{N_t} \sum_{t=1}^{N_t} \boldsymbol{\phi}(i, j, t) = \mathbf{0}. \tag{4}$$

**GLR and LLR dissimilarity metrics**

To register all images in $\tilde{f}(\mathbf{x}, t)$, we can use either the previously described GLR metric [18, 19] or the proposed LLR metric. To use the GLR metric, one firstly reformulates the entire series

of images into a *Casorati matrix*:

$$\mathbf{C} = \begin{bmatrix} \tilde{f}(\mathbf{x}_1, 1) & \tilde{f}(\mathbf{x}_1, 2) & \cdots & \tilde{f}(\mathbf{x}_1, N_t) \\ \tilde{f}(\mathbf{x}_2, 1) & \tilde{f}(\mathbf{x}_2, 2) & \cdots & \tilde{f}(\mathbf{x}_2, N_t) \\ \vdots & \vdots & \ddots & \vdots \\ \tilde{f}(\mathbf{x}_L, 1) & \tilde{f}(\mathbf{x}_L, 2) & \cdots & \tilde{f}(\mathbf{x}_L, N_t) \end{bmatrix} \in \mathbb{R}^{L \times N_t}, \quad (5)$$

where $L = N_x N_y$ is the total number of voxels, and the l-th row holds the signal in the l-th voxel. GLR assumes the Casorati matrix is rank-deficient if all images are well aligned. The GLR dissimilarity is thus determined by the nuclear norm of $\mathbf{C}$:

$$D_{GLR} = \|\mathbf{C}\|_* = \sum_{t=1}^{N_t} \sigma_t, \quad (6)$$

where $\sigma_t$ is the t-th singular value of $\mathbf{C}$.

In cardiac cine imaging, however, signals may be spatially inconsistent even after registration, due to a set of confounders such as motion, blood flow, and artifacts. As a result, the post-registration global rank may be still high, rendering GLR less effective for the guidance of the registration. Figure 1 illustrates this problem using a realistic cine movie example. From the illustration, it is clear that the local signal rank is more sensitive to the elimination of motion than the global signal rank. For example, the rank for voxels surrounding the myocardium border is reduced significantly by the registration. Therefore, we propose to use the LLR metric, which uniformly partitions the spatial domain into $N_c$ overlapped square patches with a prespecified patch size and spacing, reformulates the patches into local Casorati matrices, and computes the sum of the nuclear norm of these local Casorati matrices:

$$D_{LLR} = \sum_{\kappa=1}^{N_c} \|\mathbf{C}_\kappa\|_* = \sum_{\kappa=1}^{N_c} \sum_{t=1}^{N_t} \sigma_{\kappa,t}, \tag{7}$$

where $\mathbf{C}_\kappa$ is the local Casorati matrix associated with the κ-th patch, and $\sigma_{\kappa,t}$ is the t-th singular value of $\mathbf{C}_\kappa$.

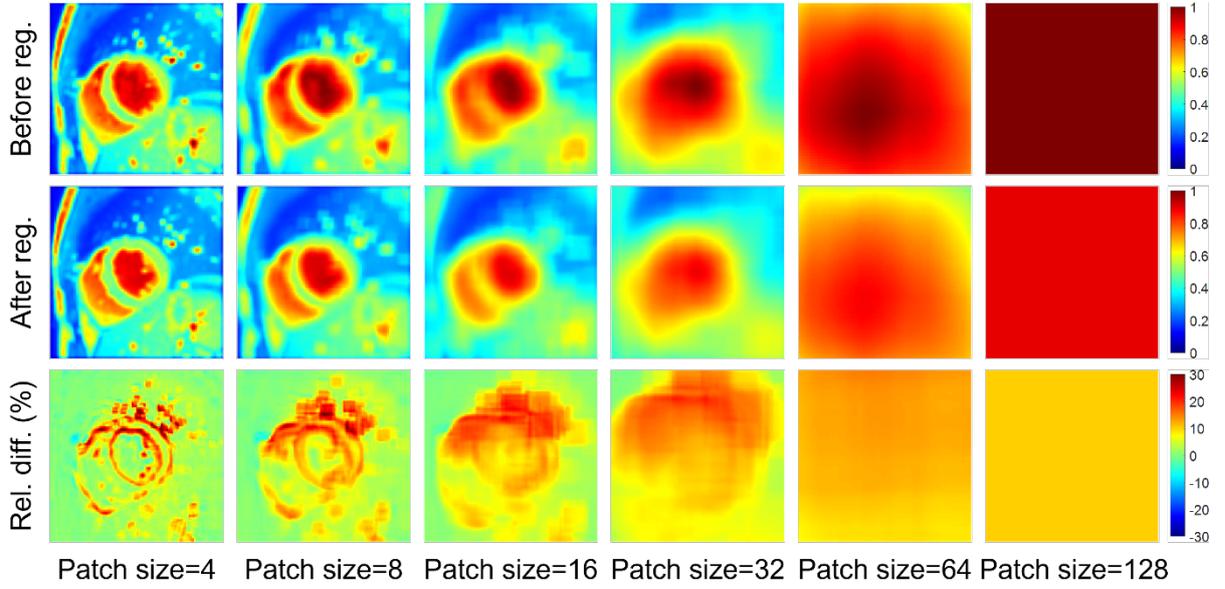

**Figure 1.** The maps of LLR cost—a surrogate of the patchwise signal rank—of a cine image sequence in the ACDC dataset before (Row 1) and after (Row 2) registration. Row 3 shows the relative difference between Row 1 and Row 2. Note that the rank reduction is highly region-dependent, with high reductions (~40%) in regions with large motion, such as the myocardial borders. Some regions have a high rank even after image registration, such as the blood pool. GLR, which is equivalent to LLR with a patch size of 128 (the rightmost column), assumes the rank is globally low after image registration. However, as the rightmost column shows, the post-registration global rank is still high, resulting in only a ~10% rank reduction.

**Regularization**

To enforce the spatiotemporal regularity of the deformation field, we use a combination of spatial and temporal regularization terms. For the spatial smoothness, we introduce a bending-

energy-based regularization term over the spatial domain [24]:

$$R_{spatial} = \sum_{t=1}^{N_t} \sum_{x \in \Omega} \left\| \frac{\partial^2 \mathbf{T}}{\partial x^2}(\mathbf{x}, t) \right\|_2^2 + 2 \left\| \frac{\partial^2 \mathbf{T}}{\partial x \partial y}(\mathbf{x}, t) \right\|_2^2 + \left\| \frac{\partial^2 \mathbf{T}}{\partial y^2}(\mathbf{x}, t) \right\|_2^2. \tag{8}$$

For the temporal smoothness, we introduce a second-order regularization term over the temporal domain:

$$R_{temporal} = \sum_{t=1}^{N_t} \sum_{x \in \Omega} \left\| \frac{\partial^2 \mathbf{T}}{\partial t^2}(\mathbf{x}, t) \right\|_2^2. \tag{9}$$

We approximate the partial derivates in $R_{spatial}$ and $R_{temporal}$ by the finite differences. Furthermore, given the cyclic motion of the heart, we adopt cyclic finite differences in $R_{temporal}$ to enforce the consistency of the motion between the first frame and the last frame.

As a result, the proposed LLR-based groupwise registration (Groupwise-LLR) is formulated as

$$\min_{\phi} D_{LLR} + \lambda R_{spatial} + \mu R_{temporal}$$

$$\text{s.t.s} \frac{1}{N_t} \sum_{t=1}^{N_t} \phi(i, j, t) = \mathbf{0}, \tag{10}$$

where $\lambda$ and $\mu$ are the regularization coefficients, and the constraint is from Eq 4. Inspired by a previous study on groupwise registration [25], we implemented the projected gradient descent algorithm to solve this problem in a coarse-to-fine multi-resolution framework. In this framework, the proposed method firstly estimates the deformation at the lowest resolution level to roughly align the images. Then, the method gradually refines the deformation at higher resolution levels to achieve more accurate alignment. We initialize the control point mesh $\boldsymbol{\phi}(i, j, t)$ as zero, leading to zero displacements across all cardiac phases.

**Strain estimation**

After obtaining the temporally resolved transformation field, we estimate the transformation field from the first image (end-diastole) to each later image by a two-step process, in which the coordinate $\mathbf{x}$ in the end-diastolic image is firstly mapped to the common reference frame, and then to the t-th image. Mathematically, this process is represented by

$$\mathbf{T}_{1 \to t}(\mathbf{x}) = \mathbf{T}(\mathbf{T}^{-1}(\mathbf{x}, 1), t), \tag{11}$$

where $\mathbf{T}_{1 \to t}(\mathbf{x})$ represents the transformation field from the first image to the t-th image. Based on the estimated $\mathbf{T}_{1 \to t}(\mathbf{x})$, the Green-Lagrange strain tensor [26] is evaluated by

$$\mathbf{E}(\mathbf{x}, t) = \frac{1}{2}\left[\left(\nabla_{\mathbf{x}} \mathbf{T}_{1 \to t}(\mathbf{x})\right)^{\mathrm{T}} \nabla_{\mathbf{x}} \mathbf{T}_{1 \to t}(\mathbf{x}) - \mathbf{I}\right], \tag{12}$$

where $\nabla_{\mathbf{x}}$ represents the Jacobian operator about $\mathbf{x}$ (implemented using finite differences) and $\mathbf{I}$ is an identity matrix of size two. The strain along a certain direction is then computed by

$$e_{\mathbf{u}}(\mathbf{x}, t) = \left(\mathbf{u}(\mathbf{x})\right)^{\mathrm{T}} \mathbf{E}(\mathbf{x}, t) \mathbf{u}(\mathbf{x}), \tag{13}$$

where $\mathbf{u}(\mathbf{x})$ is the strain direction field determined by segmentation of the left ventricular myocardium in the end-diastolic image. For those long-axis slices, $\mathbf{u}(\mathbf{x})$ represents the longitudinal direction; for those short-axis slices, $\mathbf{u}(\mathbf{x})$ represents the circumferential or radial direction. The global longitudinal strain (GLS) is estimated by averaging the strain values over the left ventricular myocardium within a single long-axis slice. The global circumferential strain (GCS) and global radial strain (GRS) are estimated by averaging the strain values over the left ventricular myocardium within a specified slice (e.g., base, mid-ventricle, apex) or across multiple slices covering the entire left ventricle.

**Baseline methods**

The proposed Groupwise-LLR was compared with the Farneback optical flow [27], a commercially available pairwise registration method [28], and the GLR-based groupwise registration (Groupwise-GLR) based on both simulated and in-vivo data. Among these alternative CMR-FT methods, the optical flow and pairwise registration estimate the transformation field between each pair of the neighboring images, i.e., $\mathbf{T}_{t-1 \to t}(\mathbf{x})$, $2 \leq t \leq N_t$. The transformation field $\mathbf{T}_{1 \to t}(\mathbf{x})$ is then a composition of the neighboring transformation fields. Groupwise-GLR was implemented in the same way as Groupwise-LLR except for the use of a GLR dissimilarity metric. The proposed method was also compared with two general groupwise registration methods, nD+t B-Splines [16] and pTVreg [21], in simulations and in vivo tracking. Among them, nD+t B-Splines uses an nD+t B-spline deformation model and a sum of the voxelwise signal variances as the dissimilarity metric. pTVreg uses a similar LLR metric as the proposed one. However, pTVreg extracts nonoverlapped patches to compute the LLR metric and fixes the patch size for all resolution levels. In comparison, Groupwise-LLR partitions the spatial domain into overlapped patches to enhance the continuity of the estimated motion among neighboring patches, and adjusts the patch size along with the image resolution to improve the consistency of the multi-resolution optimization. Additionally, pTVreg is characterized by the total variation regularization, which is different from the second-order

Tikhonov regularization used in this work. Furthermore, it should be emphasized that the proposed Groupwise-LLR is the first strain estimation method based on groupwise registration while the two baseline groupwise registration methods have not been applied for strain estimation before.

**Implementation details**

All methods were implemented with MATLAB (R2022a, MathWorks, Natick, MA, USA). The parameters of each method were manually tuned for the simulated and in vivo experiments. For the proposed Groupwise-LLR, the number of resolution levels was 3 and the downsampling factor between two adjacent resolution levels was 2. The regularization coefficients were not reweighted between different resolution levels. The cubic interpolation was used for image downsampling and the linear interpolation was used for image warping. The patch size and spacing used to sample the local Casorati matrices were 5 and 3 pixels at the lowest resolution level and doubled each time moving to the next level. The control point spacing, spatial regularization coefficient $\lambda$, and temporal regularization coefficient $\mu$ were 6 and 7 pixels, $6\times10^{-4}$ and $1\times10^{-3}$, 0.06 and 0.1 for the simulated dataset and two in vivo datasets, respectively. We determined the regularization coefficients using a semi-quantitative approach based on small separate trial datasets. For the simulated experiments, we firstly set $\lambda = 5\times10^{-4}$ based on a manual inspection of the spatial smoothness of the displacement fields. We then varied the ratio $\mu/\lambda$ from 0.1 to 1000, and found that $\mu/\lambda = 100$ minimized the EPE over the simulated trial dataset. Thus, we fixed $\mu/\lambda = 100$ hereafter. We varied $\lambda$ from $2\times10^{-4}$ to $1\times10^{-3}$ with a step of $2\times10^{-4}$, and found that $\lambda = 6\times10^{-4}$ minimized the EPE over the simulated trial dataset. For the in-vivo experiments, we tried different values of $\lambda$ from $5\times10^{-4}$ to $2.5\times10^{-3}$ with a step of $5\times10^{-4}$, and found that $\lambda = 1\times10^{-3}$ minimized the mean contour distances over the in-vivo trial dataset. The optimization at each resolution level was terminated when the relative value change of the cost function was less than a tolerance of $1\times 10^{-5}$.

**Simulations**

The simulated cine MRI dataset was generated by a digital phantom (XCAT) [22] and its MRI extension MRXCAT [29]. Specifically, XCAT was used to generate the cardiac-phase-resolved 3D tissue masks of 20 subjects, each with a different gender, cardiac anatomy, and cardiac motion pattern. The corresponding transformation field was also generated for each subject. Interpolation was then used to generate 2D cine movies in a 2-chamber slice and 7 or 8 short-axis slices, with a 1.5 mm×1.5 mm resolution and 24 frames. MRXCAT was then used to generate the signal intensity of each tissue in cine MRI. The studied feature tracking methods were performed in each slice to estimate the strains. The tracking quality was evaluated by end-point error (EPE), which is the difference between the estimated transformation field and its ground truth, averaged over the entire myocardium at end-systole or over the entire cardiac cycle. Strain error was computed at end-systole for both voxelwise (VSE) and global (GSE) levels, where the former was an average of absolute strain difference between the estimated strain and ground truth over the entire myocardium, and the latter was the absolute difference between the global estimated strain and global ground truth strain.

**Experiments on the ACDC dataset**

The ACDC dataset [23] contains 100 subjects evenly distributed over five study groups: dilated cardiomyopathy (DCM), myocardial infarction (MI), abnormal right ventricle (ARV), hypertrophic cardiomyopathy (HCM), and normal cardiac anatomy and function (NOR). For each subject, a series of short-axis cine images were acquired with a balanced steady-state free-precession (bSSFP) sequence on a 1.5T (Siemens Area, Siemens Medical Solutions, Germany) or 3T (Siemens Trio Tim, Siemens Medical Solutions, Germany) clinical scanner. Manual segmentation of the end-diastolic and end-systolic myocardium is provided by the dataset. To evaluate the tracking accuracy of a method, the end-diastolic epicardial and endocardial contours were first tracked to the end-systolic phase. Then, the mean distance between the tracked and annotated end-systolic contours was computed in the basal, mid-ventricular, and apical slices. To evaluate the inter-observer reproducibility of the measured strains, the end-diastolic endocardium and epicardium were independently contoured by a reader (3 years of cardiac MR experience) in a subset (n = 30) of the ACDC dataset, and the intraclass correlation coefficients (ICC) of the end-systolic global strains (radial and circumferential) based on the annotations by the dataset and the reader were calculated. To evaluate the intra-observer reproducibility, the same set of data was annotated by the same reader over one year later, and the ICC based on the two annotations by this reader were calculated.

**Experiments on the patient dataset**

The patient study was approved by the institutional review board. All patients provided informed written consent. The clinical dataset contains 16 patients (9 male, age 51 ± 14 years) scanned between February 2016 and February 2017 on a 3T scanner (Philips Ingenia Elition, Philips Healthcare, Nederland). Clinical indications included normal (n = 3), coronary heart diseases (n = 4), connective tissue diseases (n = 4), nonischemic heart diseases (n = 3), and other (n = 2). All patients were scanned with a bSSFP cine sequence and a tagging-MRI sequence. The bSSFP cine sequence used the following parameters: TR = 2.8 ms, TE = 1.4 ms, bandwidth = 1890 Hz/pixel, flip angle = 45°, FOV = 268×280 mm$^2$, matrix size = 134×140, slice thickness = 7 mm, number of phases = 30. The tagging-MRI sequence had the following parameters: TR = 5.9 ms, TE = 3.6 ms, bandwidth = 432 Hz/pixel, flip angle = 10°, tag spacing = 7 mm, FOV = 229×280 mm$^2$, matrix size = 154×188, slice thickness = 8 mm, number of phases = 9 to16. For each subject, only the matched cine and tagging images at basal, mid-ventricular, and apical short-axis slices were analyzed. The left ventricular myocardium in each end-diastolic cine or tagging image was annotated according to the AHA 16-segment model [30]. The tagging images were analyzed by pairwise registration following a previous work [31]. The end-systolic global and segmental strains estimated by the CMR-FT methods were compared with those by tagging using ICC.

**Statistics**

Continuous variables were reported as mean ± standard deviation. The Wilcoxon signed rank test was performed to determine whether two continuous variables were equivalent. The effect size of the Wilcoxon signed rank test was computed by dividing the absolute standardized test statistic z by the square root of the number of sample pairs. Following Cohen's classification [32], the threshold of the effect size for a small, medium, and large effect are 0.1, 0.3, and 0.5, respectively. For evaluation of the inter/intra-observer reproducibility, ICC was calculated using a single rating, absolute agreement, and 2-way mixed-effects model; for evaluation of the cine-tagging correlation, ICC was calculated using a single rating, consistency, and 2-way mixed-effects model [33]. Bland-Altman analysis was also performed to evaluate the inter/intra-observer reproducibility of each method. All statistical analyses were performed with MATLAB and SPSS (Version 26, IBM, New York, USA). P values less than 0.05 were considered statistically significant.

# Results

**Simulations**

Figure 2 shows changes of the EPE, VSE, and global strains over the entire cardiac cycle for one simulated subject. Groupwise-LLR achieved the lowest errors for tracking and strain estimation over the entire cardiac cycle. The largest improvement of Groupwise-LLR was found at end-systole (orange arrows). In late-diastole (blue arrows), the four groupwise methods were more accurate than the pairwise registration and Farneback optical flow, both of which suffered from the drift effect. Groupwise-LLR led to improved strain accuracies compared with the other methods, with a substantial improvement in the radial strain estimation.

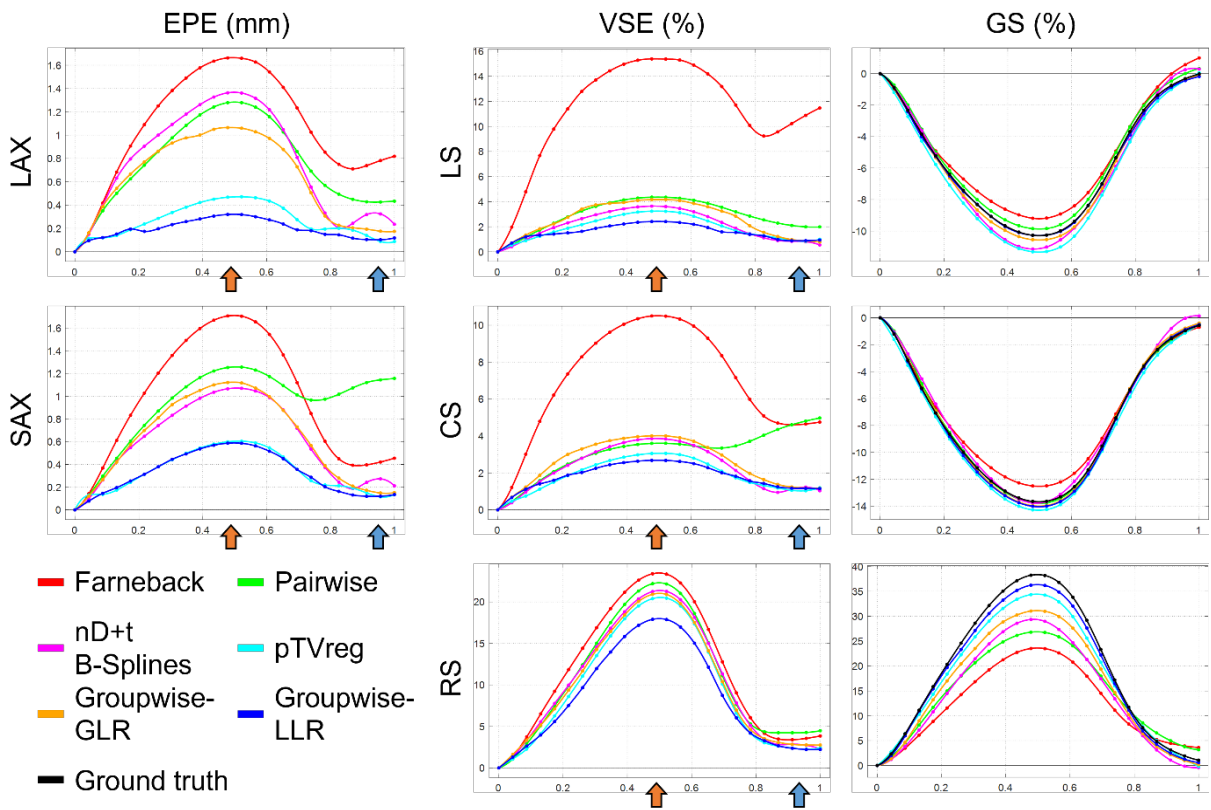

**Figure 2.** Plots of the end-point errors (EPE), voxelwise strain errors (VSE), and global strains (GS) over a cardiac cycle for one simulated subject. The curves were plotted as a function of normalized time. Groupwise-LLR achieved the lowest errors among the studied methods for tracking, voxelwise strain estimation, and global strain estimation. The largest improvement occurred at end-systole (orange arrows). In late-diastole, the four groupwise methods were more accurate than Farneback optical flow and pairwise registration (blue arrows) due to the elimination of the drift effect. LAX = long-axis; SAX = short-axis; LS = longitudinal strain; CS = circumferential strain; RS = radial strain.

Table 1 compares the values of $EPE_{ES}$, $EPE_{all}$, $VSE_{ES}$, and $GSE_{ES}$ from the studied methods over the entire simulated dataset. Groupwise-LLR achieved lower $EPE_{ES}$ and $EPE_{all}$ compared with the other methods for both long-axis and short-axis views (almost all $p < 0.05$). Furthermore, Groupwise-LLR achieved lower $VSE_{ES}$ and $GSE_{ES}$ than the other methods for all strain types (almost all $p < 0.05$). In particular, the improvement of radial strain accuracy by Groupwise-LLR was relatively substantial.

Table 1. Comparison of the End-Point Errors, Voxelwise Strain Errors, and Global Strain Errors Between Different Methods on the Simulated Dataset

| Method | $EPE_{ES}$ (mm) | | $EPE_{all}$ (mm) | | $VSE_{ES}$ (%) | | | $GSE_{ES}$ (%) | | |
|---|---|---|---|---|---|---|---|---|---|---|
| | LAX | SAX | LAX | SAX | LS | CS | RS | LS | CS | RS |
| Farneback | 1.36 ± 0.47* es = 0.87 | 1.46 ± 0.29* es = 0.87 | 0.82 ± 0.29* es = 0.87 | 0.80 ± 0.18* es = 0.87 | 10.34 ± 3.52* es = 0.87 | 9.17 ± 1.49* es = 0.87 | 19.19 ± 5.24* es = 0.87 | 1.13 ± 0.74* es = 0.86 | 1.46 ± 0.81* es = 0.87 | 8.93 ± 5.44* es = 0.86 |
| Pairwise | 1.07 ± 0.45* es = 0.86 | 0.98 ± 0.17* es = 0.87 | 0.64 ± 0.22* es = 0.87 | 0.68 ± 0.13* es = 0.87 | 3.30 ± 0.95* es = 0.65 | 2.97 ± 0.48* es = 0.76 | 16.01 ± 4.53* es = 0.87 | 0.75 ± 0.45* es = 0.84 | 0.53 ± 0.33* es = 0.79 | 6.61 ± 4.04* es = 0.82 |
| nD+t B-Splines | 1.02 ± 0.46* es = 0.87 | 0.96 ± 0.19* es = 0.87 | 0.57 ± 0.24* es = 0.87 | 0.54 ± 0.12* es = 0.87 | 3.86 ± 0.91* es = 0.87 | 3.65 ± 0.65* es = 0.87 | 16.20 ± 4.36* es = 0.87 | 0.36 ± 0.34* es = 0.51 | 0.59 ± 0.42* es = 0.76 | 6.22 ± 4.35* es = 0.76 |
| pTVreg | 0.55 ± 0.36* es = 0.38 | 0.52 ± 0.09* es = 0.61 | 0.29 ± 0.17 es = 0.25 | 0.29 ± 0.06* es = 0.52 | 3.42 ± 1.89* es = 0.50 | 2.76 ± 0.27* es = 0.54 | 14.59 ± 4.10* es = 0.87 | 0.29 ± 0.30 es = 0.36 | 0.22 ± 0.27* es = 0.38 | 2.24 ± 1.97* es = 0.45 |
| Groupwise-GLR | 0.89 ± 0.33* es = 0.87 | 0.97 ± 0.21* es = 0.87 | 0.49 ± 0.16* es = 0.87 | 0.54 ± 0.13* es = 0.87 | 3.45 ± 0.91* es = 0.86 | 3.66 ± 0.45* es = 0.87 | 15.83 ± 4.24* es = 0.87 | 0.33 ± 0.17* es = 0.70 | 0.46 ± 0.34* es = 0.75 | 4.28 ± 3.25* es = 0.72 |
| Groupwise-LLR | **0.38 ± 0.12** | **0.46 ± 0.12** | **0.23 ± 0.07** | **0.26 ± 0.07** | **2.25 ± 0.52** | **2.54 ± 0.45** | **12.86 ± 3.81** | **0.18 ± 0.15** | **0.13 ± 0.10** | **1.57 ± 1.11** |

Bold: The lowest error in that column. *: Significantly higher error than Groupwise-LLR based on the one-tailed Wilcoxon signed rank test. EPE = end-point error; VSE = voxelwise strain error; GSE = global strain error; ES = end-systolic; LAX = long-axis; SAX = short-axis; LS = longitudinal strain; CS = circumferential strain;

RS = radial strain; es = effect size.

**ACDC dataset**

*Tracking quality*

Figure 3 compares the tracking quality of all the studied methods, measured by the mean distance between each predicted contour and manually drawn contour at end-systole. For both epicardial and endocardial contours, Groupwise-LLR achieved significantly lower mean distances (1.72 ± 0.57 mm and 2.11 ± 1.33 mm) than Farneback optical flow (1.88 ± 0.72 mm and 2.50 ± 1.61 mm, both p < 0.01), pairwise registration (1.80 ± 0.72 mm and 2.30 ± 1.47 mm, both p < 0.01), nD+t B-Splines (1.87 ± 0.77 mm and 2.21 ± 1.51 mm, both p < 0.01), pTVreg (1.80 ± 0.69 mm and 2.29 ± 1.56 mm, both p < 0.01), and Groupwise-GLR (1.85 ± 0.73 mm, p < 0.01 and 2.17 ± 1.48 mm, p < 0.05). These results demonstrate the superiority of Groupwise-LLR in terms of tracking quality.

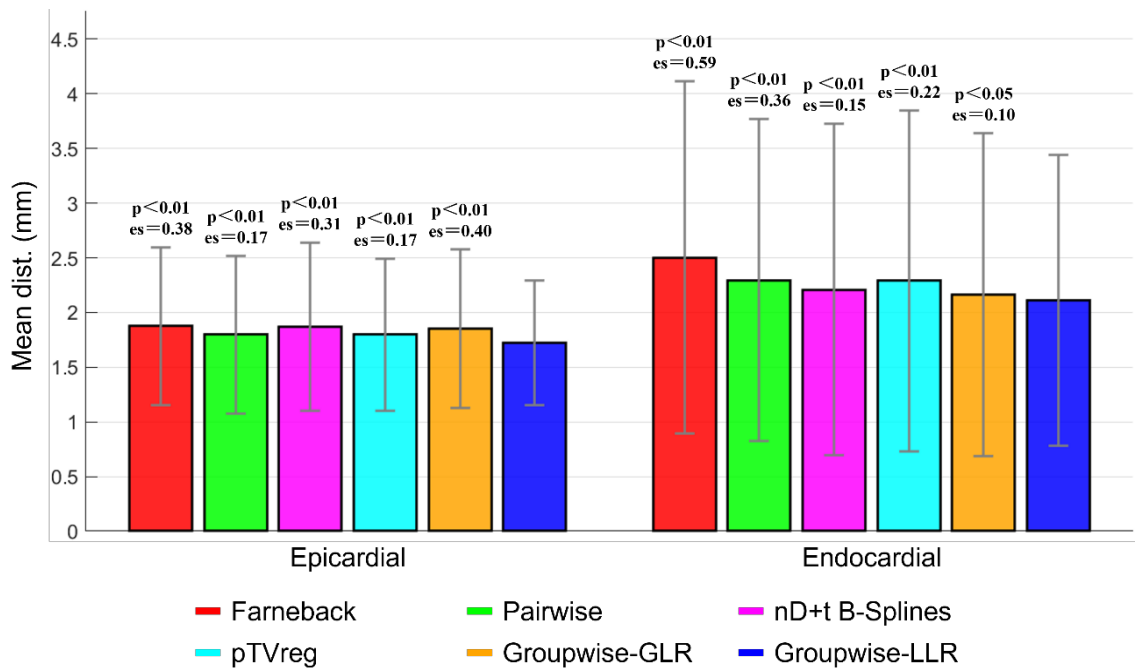

**Figure 3.** Comparison of the tracking quality of different methods on the ACDC dataset. The height of each bar represents the mean value of the mean distance between the predicted contour by each method and the manually drawn contour at end-systole. The error bar indicates the standard deviation. For each subject, only the basal, mid-ventricular, and apical slices were used for the contour tracking. The p value and effect size (es) from the one-sided Wilcoxon sign rank test are labeled over each bar for the three alternative methods. Groupwise-LLR showed a significant reduction of mean distance compared with the other methods for both the epicardial and endocardial contours.

*Strain accuracy*

Figure 4 shows the global strain curves of three subjects, each from a different study group. For every CMR-FT method, the MI and DCM subjects had lower strains than the control. However, whereas the two groupwise methods generated strains closer to zero at late-diastole, the two pairwise methods generated nonzero strains in late-diastole due to the drift effect (arrows). Figure 5 shows the mid-ventricular strain maps of an MI subject, who has wall thinning and akinesia in the septum. All four methods captured the reduction of radial strain in the septum; however, Farneback showed more inaccurate strains compared with the other methods. The late-diastolic displacement and strain estimates also appeared more accurate with the groupwise registration methods, due to the elimination of the drift effect. An additional movie file shows the estimated strain maps across the whole cardiac cycle [see Additional file 1].

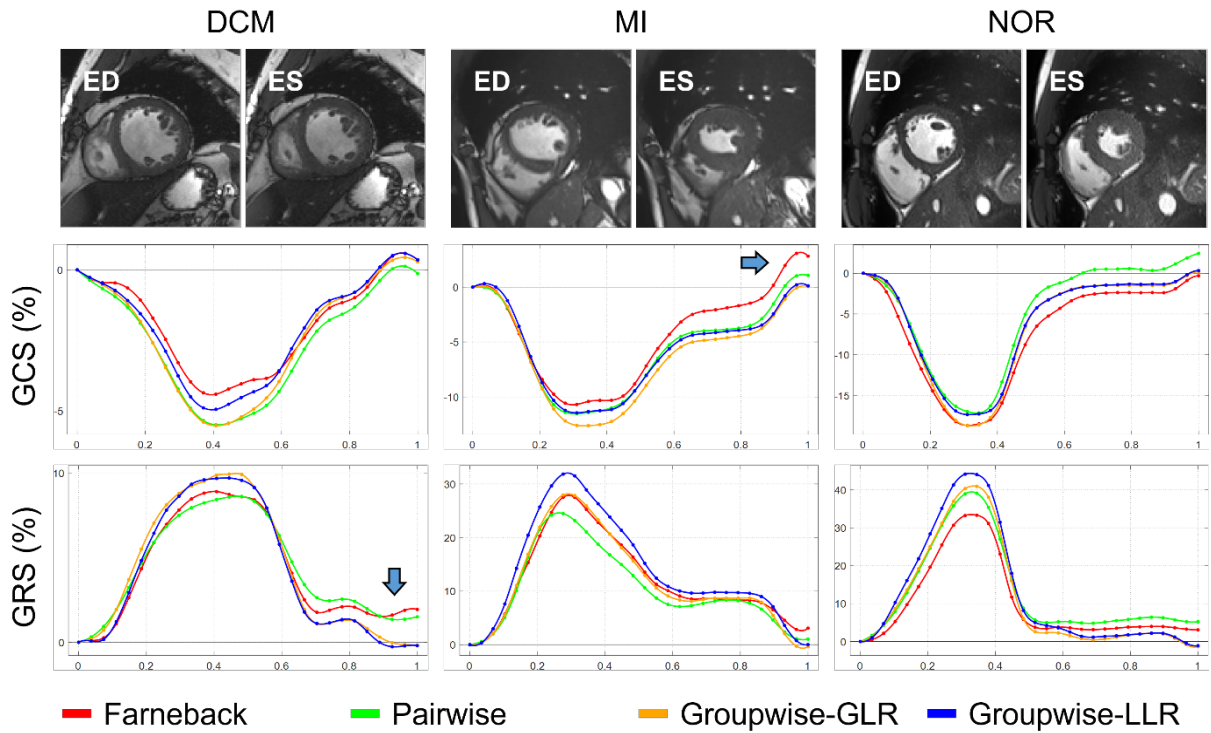

**Figure 4.** Global circumferential and radial strain curves of three subjects from different study groups in the ACDC dataset. The curves were plotted as a function of normalized time. The end-diastolic and end-systolic mid-ventricular images of each subject are shown at the top of each column. Normally, the strain at the end of the cardiac cycle should be nearly zero due to the periodicity of the cardiac motion. However, in the presence of the drift effect, the estimated late-diastolic strain can be significantly different from zero, which is observed for the Farneback optical flow and pairwise registration but not for the groupwise methods. ED = end-diastolic; ES = end-systolic; DCM = dilated cardiomyopathy; MI = myocardial infarction; NOR = normal cardiac anatomy and function; GCS = global circumferential strain; GRS = global radial strain.

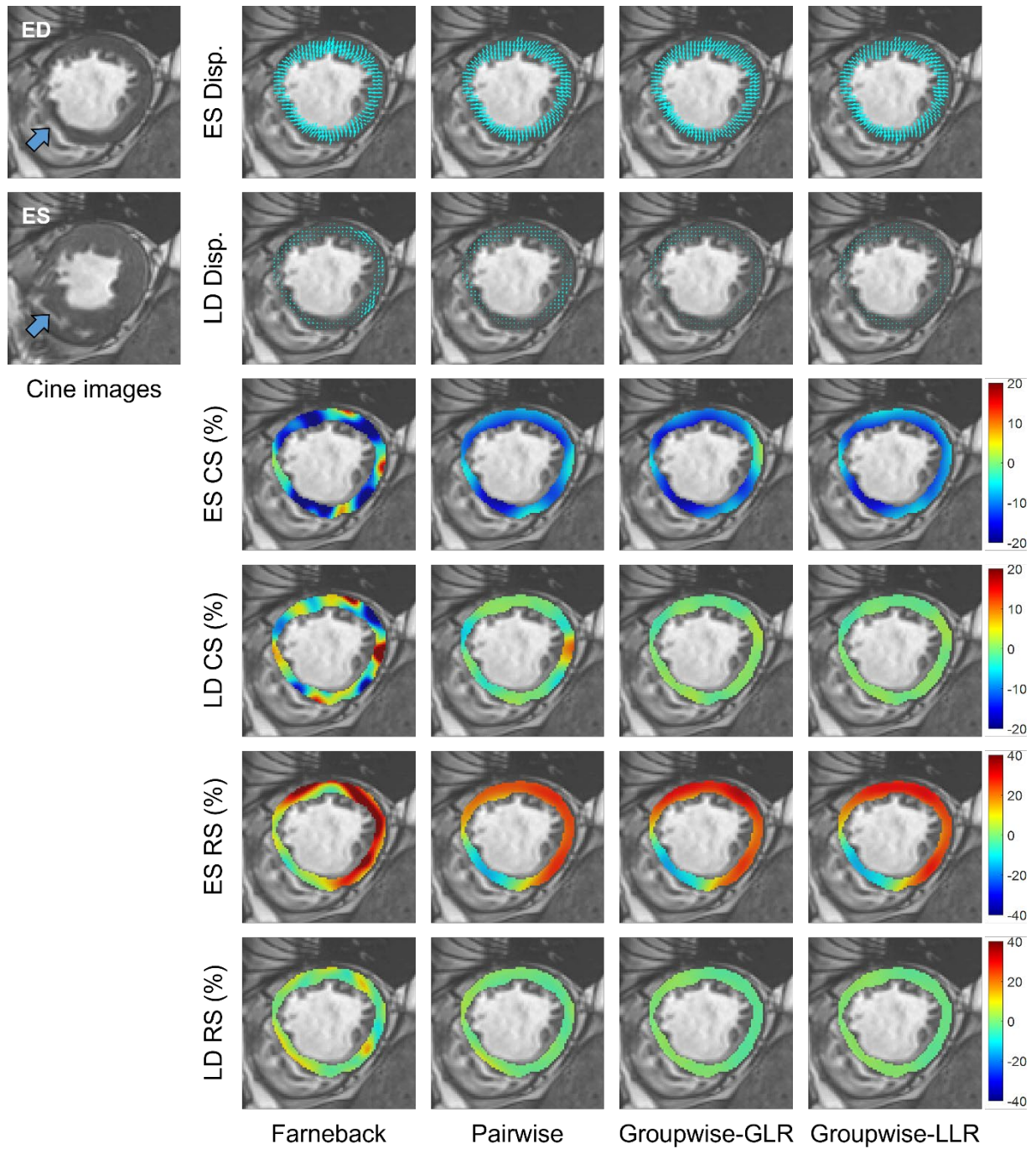

**Figure 5.** Mid-ventricular cine images of a subject with myocardial infarction in the ACDC dataset and the corresponding end-systolic and late-diastolic displacement fields, circumferential strain maps, and radial strain maps. The septal wall (arrows) exhibited abnormal thinning and akinesia due to the presence of infarction. Farneback caused nonsmooth estimates of the circumferential strain and underestimates of the radial strain in the anterior wall. Both Farneback and pairwise registration suffered from the drift effect and manifested nonnegligible displacements and strains in the late-diastolic phase. The groupwise registration

methods demonstrated a smoother circumferential strain map, more accurate radial strain estimates, and a reduction of drift effect. ED = end-diastolic; ES = end-systolic; LD = late-diastolic (corresponding to the last frame); CS = circumferential strain; RS = radial strain.

Figure 6 shows the mean and 95% confidence interval of the global strain curves over each study group obtained by different methods. One can notice the systematic occurrence of drift effects in late-diastole with the two pairwise methods (arrows), which was largely reduced by the two groupwise methods. The mean absolute value of the late-diastolic GCS/GRS were 1.51%/4.11% for Farneback optical flow and 1.33%/4.09% for the pairwise registration. In comparison, the mean absolute value of the late-diastolic GCS/GRS were 0.74%/1.60% for Groupwise-GLR, and 1.00%/1.98% for Groupwise-LLR.

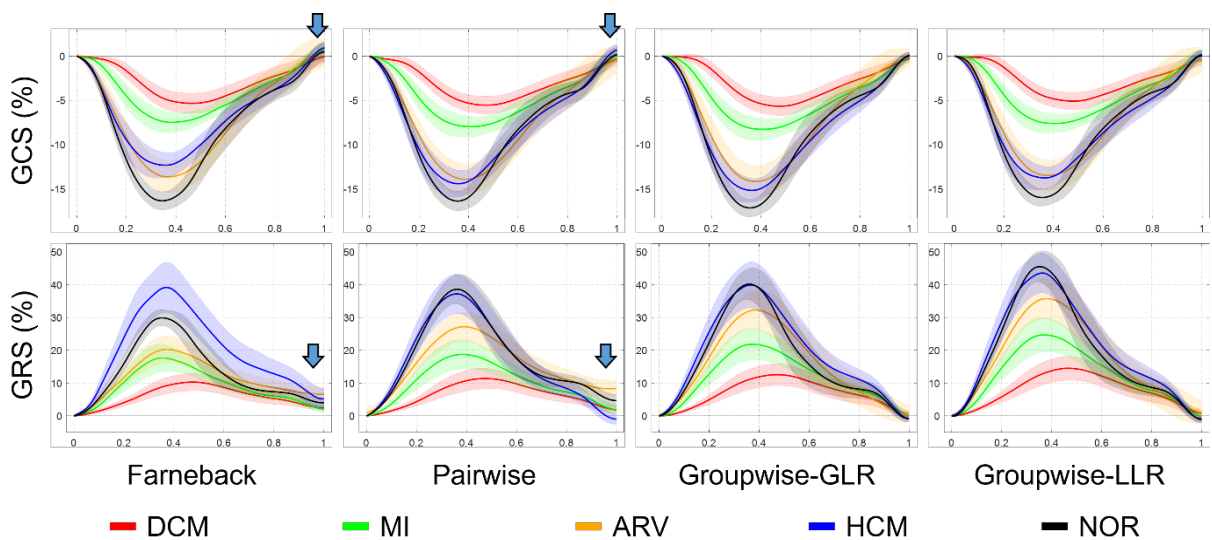

**Figure 6.** The mean and confidence interval of the global circumferential and radial strain curves over each study group (n=20) in the ACDC dataset. Each curve represents an average global strain, with different colors representing different study groups. The shade around each curve represents the 95% confidence interval at each time point. The arrows highlight the systematic drift effect of the Farneback optical flow and pairwise registration. GCS = global circumferential strain; GRS = global radial strain; DCM = dilated cardiomyopathy; MI = myocardial infarction; ARV = abnormal right ventricle; HCM = hypertrophic cardiomyopathy; NOR = normal cardiac anatomy and function.

Table 2 shows the end-systolic global strains obtained by different methods for each study group. Groupwise-LLR generated a significantly greater GRS for every study group than the other methods, while the GCS was more similar. These trends were consistent with the findings from the simulated dataset.

**Table 2. Comparison of the End-Systolic Global Circumferential and Radial Strains Between Different Methods on the ACDC Dataset**

| Method | ES GCS (%) | | | | | ES GRS (%) | | | | |
|---|---|---|---|---|---|---|---|---|---|---|
| | DCM | MI | ARV | HCM | NOR | DCM | MI | ARV | HCM | NOR |
| Farneback | -5.5 ± 2.6<br>es = 0.38 | -7.8 ± 2.3<br>es = 0.27 | -14.2 ± 3.8<br>es = 0.03 | -12.4 ± 3.1*<br>es = 0.61 | -16.9 ± 2.1<br>es = 0.29 | 10.0 ± 5.4*<br>es = 0.88 | 19.5 ± 8.3*<br>es = 0.73 | 22.1 ± 7.8*<br>es = 0.88 | 39.8 ± 16.8*<br>es = 0.45 | 32.1 ± 6.0*<br>es = 0.88 |
| Pairwise | -5.6 ± 2.2*<br>es = 0.75 | -8.3 ± 2.5*<br>es = 0.46 | -14.5 ± 3.4<br>es = 0.36 | -14.5 ± 3.3*<br>es = 0.48 | -16.8 ± 2.3<br>es = 0.26 | 11.2 ± 6.5*<br>es = 0.81 | 20.4 ± 8.9*<br>es = 0.85 | 29.7 ± 8.2*<br>es = 0.82 | 37.8 ± 13.2*<br>es = 0.72 | 40.3 ± 10.2*<br>es = 0.79 |
| Groupwise-GLR | -5.7 ± 2.2*<br>es = 0.77 | -8.6 ± 2.5*<br>es = 0.77 | -14.9 ± 3.7*<br>es = 0.58 | -15.3 ± 2.9*<br>es = 0.83 | -17.6 ± 1.9*<br>es = 0.82 | 12.5 ± 7.0*<br>es = 0.77 | 24.2 ± 9.6*<br>es = 0.74 | 35.5 ± 10.8*<br>es = 0.68 | 40.4 ± 15.3*<br>es = 0.69 | 42.4 ± 10.6*<br>es = 0.75 |
| Groupwise-LLR | -5.1 ± 2.1 | -8.0 ± 2.2 | -14.1 ± 3.5 | -13.9 ± 2.8 | -16.4 ± 2.0 | 14.4 ± 7.7 | 27.5 ± 9.9 | 32.9 ± 10.9 | 44.0 ± 13.3 | 48.4 ± 11.6 |

*: Significantly different from the strain estimated by Groupwise-LLR according to the two-tailed Wilcoxon signed rank test. ES = end-systolic; GCS = global circumferential strain; GRS = global radial strain; DCM = dilated cardiomyopathy; MI = myocardial infarction; ARV = abnormal right ventricle; HCM = hypertrophic cardiomyopathy; NOR = normal cardiac anatomy and function; es = effect size.

*Reproducibility*

Figure 7 shows the inter-observer reproducibility of each CMR-FT method using ICC [33]. For all methods, the ICC of GCS was "moderate" to "excellent" while the ICC of GRS was "excellent". Figure 8 shows the results of the intra-observer reproducibility evaluation. For all methods, the ICC of both GCS and GRS were "excellent". These results suggest that the four methods have similar inter/intra-observer reproducibility.

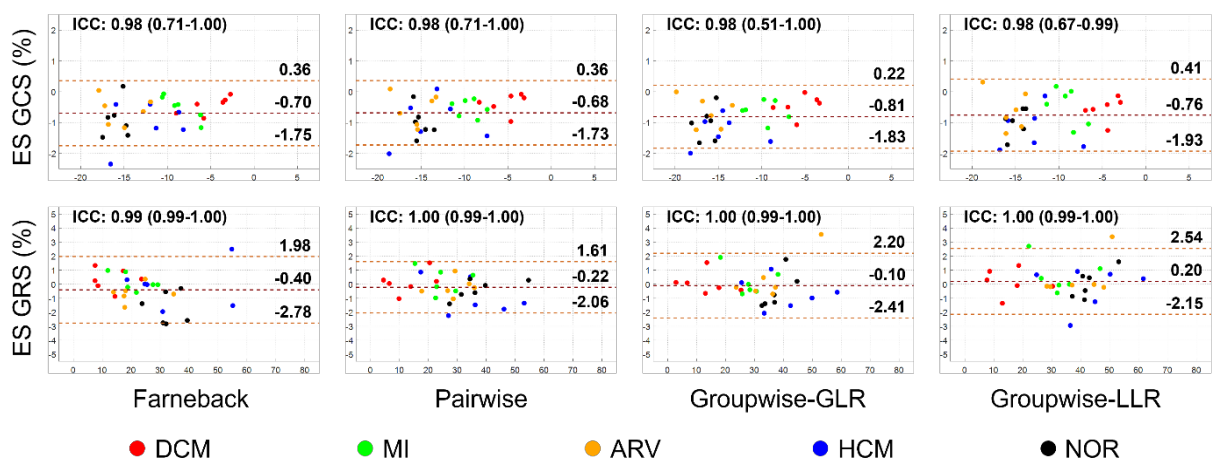

**Figure 7.** Bland-Altman analysis for the inter-observer reproducibility evaluation of the end-systolic global strains based on the ACDC dataset. The intraclass correlation coefficient (ICC) and its 95% confidence interval were labeled at the top of each plot. The four methods had similar inter-observer reproducibility. The disagreement of global circumferential strain (GCS) was greater than that of global radial strain (GRS) for every method. ES = end-systolic; DCM = dilated cardiomyopathy; MI = myocardial infarction; ARV = abnormal right ventricle; HCM = hypertrophic cardiomyopathy; NOR = normal cardiac anatomy and function.

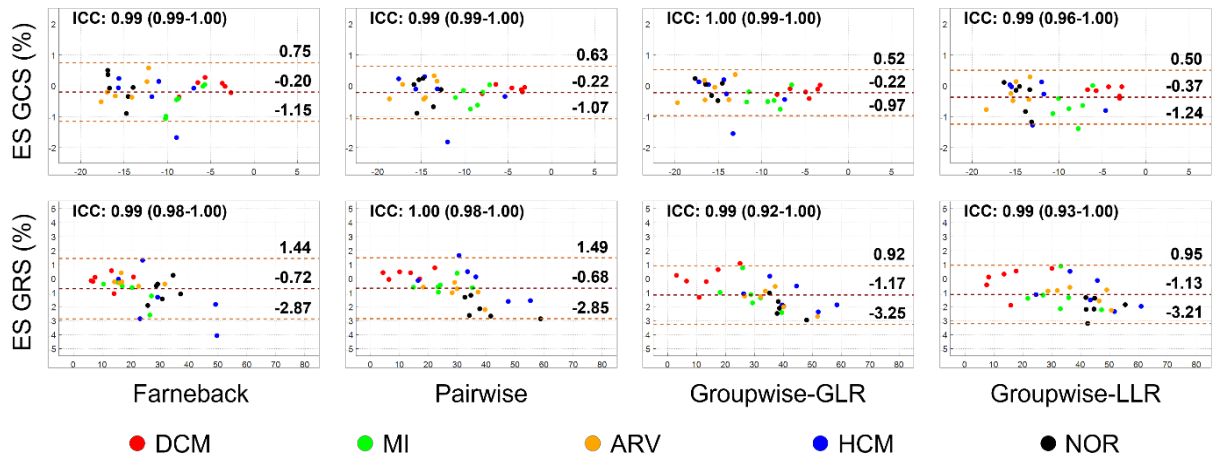

**Figure 8.** Bland-Altman analysis for the intra-observer reproducibility evaluation of the end-systolic global strains based on the ACDC dataset. The intraclass correlation coefficient (ICC) and its 95% confidence interval were labeled at the top of each plot. The four methods manifested similarly high intra-observer reproducibility. GCS = global circumferential strain; GRS = global radial strain; ES = end-systolic; DCM = dilated cardiomyopathy; MI = myocardial infarction; ARV = abnormal right ventricle; HCM = hypertrophic cardiomyopathy; NOR = normal cardiac anatomy and function.

**Patient dataset**

Table 3 compares the ICC of different methods with tagging for end-systolic global strains. For end-systolic GCS, pairwise registration, Groupwise-GLR, and Groupwise-LLR all showed a stronger correlation with tagging than Farneback optical flow. For end-systolic GRS, both groupwise methods showed a stronger correlation with tagging than the optical flow and pairwise registration in the mid-ventricular slice (r = 0.80 and 0.78 vs r = 0.55 and 0.72), apical slice (r = 0.50 and 0.57 vs r = 0.46 and 0.39), and mean over the three slices (r = 0.65 and 0.69 vs r = 0.49 and 0.56). All feature tracking-based global strains had a significant correlation with those based on tagging-MRI, except the apical radial strain measured by the pairwise registration (p = 0.06).

**Table 3. Comparison of the Correlation with Tagging between Different Methods for End-Systolic Global Strains on the Patient Dataset**

| Method | ICC for ES GCS | | | | ICC for ES GRS | | | |
|---|---|---|---|---|---|---|---|---|
| | Base | Middle | Apex | All | Base | Middle | Apex | All |
| Farneback | r = 0.64<br>p = 0.00 | r = 0.85<br>p = 0.00 | r = 0.82<br>p = 0.00 | r = 0.83<br>p = 0.00 | r = 0.53<br>p = 0.01 | r = 0.55<br>p = 0.01 | r = 0.46<br>p = 0.03 | r = 0.49<br>p = 0.02 |
| Pairwise | **r = 0.77**<br>**p = 0.00** | **r = 0.90**<br>**p = 0.00** | **r = 0.89**<br>**p = 0.00** | **r = 0.90**<br>**p = 0.00** | **r = 0.54**<br>**p = 0.01** | r = 0.72<br>p = 0.00 | r = 0.39<br>p = 0.06 | r = 0.56<br>p = 0.01 |
| Groupwise-GLR | **r = 0.77**<br>**p = 0.00** | r = 0.89<br>p = 0.00 | r = 0.83<br>p = 0.00 | r = 0.88<br>p = 0.00 | r = 0.47<br>p = 0.03 | **r = 0.80**<br>**p = 0.00** | r = 0.50<br>p = 0.02 | r = 0.65<br>p = 0.00 |
| Groupwise-LLR | **r = 0.77**<br>**p = 0.00** | r = 0.89<br>p = 0.00 | r = 0.87<br>p = 0.00 | r = 0.89<br>p = 0.00 | r = 0.53<br>p = 0.02 | r = 0.78<br>p = 0.00 | **r = 0.57**<br>**p = 0.01** | **r = 0.69**<br>**p = 0.00** |

Bold: The highest correlation coefficient(s) in that column. ICC = intraclass correlation coefficient; ES = end-systolic; GCS = global circumferential strain; GRS = global radial strain.

Figure 9 shows the bull's eye plots of the per-segment ICC of each method relative to tagging-MRI. For the circumferential strain, pairwise registration, Groupwise-GLR, and Groupwise-LLR achieved similar performances and outperformed Farneback optical flow. For the radial strain, Groupwise-LLR yielded higher correlation coefficients than Farneback optical flow, pairwise registration, and Groupwise-GLR in 13, 10, and 13 of 16 segments, respectively, suggesting an improved accuracy in regional radial strain estimates.

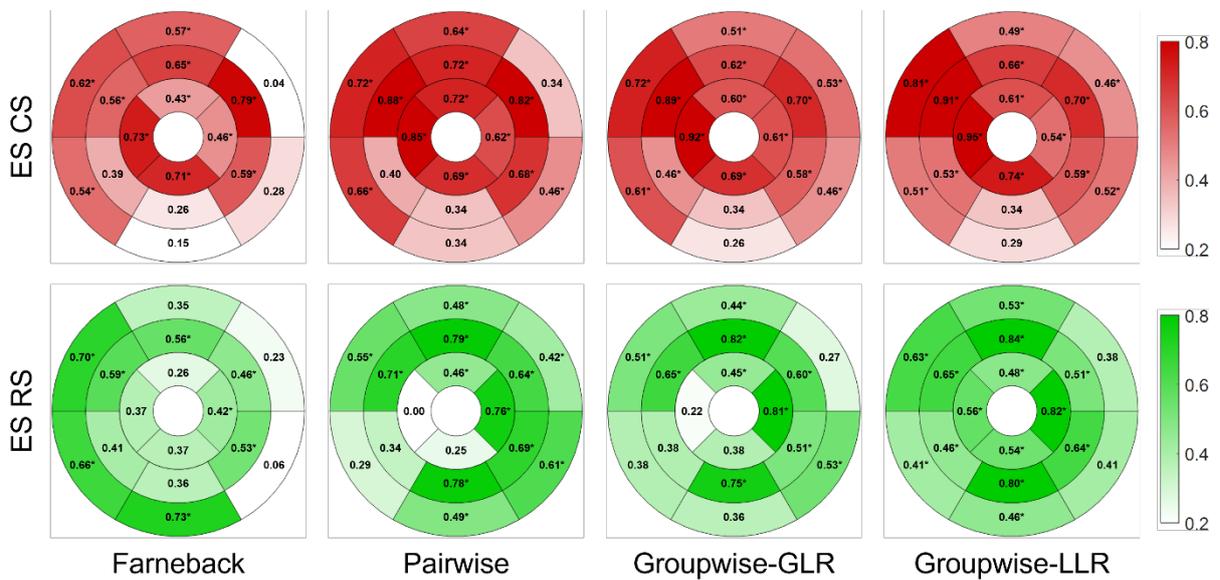

**Figure 9.** Bull's eye plots of per-segment intraclass correlation coefficients with tagging for end-systolic strains on the patient dataset. Groupwise-LLR demonstrated stronger correlations with tagging compared with the other methods, especially for the radial strain. *: Significantly correlated with the estimates by tagging-MRI. ES = end-systolic; CS = circumferential strain; RS = radial strain.

## Discussion

In this work, we propose a novel CMR-FT method based on deformable groupwise registration with an LLR dissimilarity metric. We evaluated the myocardial tracking quality, strain accuracy, and strain inter/intra-observer reproducibility of the proposed method via simulated and in vivo data. The simulations show that the proposed Groupwise-LLR method was more accurate in both tracking and strain estimation than the other methods. Results on the ACDC dataset show that the proposed method improved tracking quality and suppressed the drift effect common for optical flow and pairwise registration. However, the studied CMR-FT methods were similarly reproducible. The patient data show that Groupwise-LLR achieved a stronger correlation with tagging for the radial strains in the median slice, apical slice, and most myocardial segments.

Feature tracking based on cardiac cine MRI has drawn increasing attention due to its reproducibility, ease of use, and excellent value for risk stratification and prognosis of cardiac diseases [4, 34, 35]. Optical flow and pairwise registration are currently the major algorithms used in existing commercial software [8, 36]. However, since both of them track the myocardium frame-by-frame, small errors within each tracking step can accumulate to a larger error, resulting in a less accurate evaluation of myocardial contractility in the later cardiac phases [9]. On the other hand, although groupwise registration has been postulated to provide more accurate myocardial tracking [16, 17], little data exist to evaluate the accuracy of strains measured by the groupwise registration methods. Our findings filled this niche by showing that groupwise registration methods indeed generate more accurate strain estimates than those pairwise methods, especially for the radial strain which has been proven to be highly related to amyloidosis [37], myocarditis [38], aortic stenosis [39], etc. Furthermore, by minimizing the drift effect, the proposed method provides a more accurate estimation of the strains in the diastolic phase. The diastolic function of the heart can be affected by various diseases, such as heart failure and cardiomyopathies [40, 41]. Therefore, the improved accuracy in the estimation of diastolic strains and strain rates may translate to more accurate diagnosis and prognosis of the relevant diseases. In addition to the evaluation of the left ventricular function, the proposed method may also be used to analyze the function of other chambers, such as the right ventricle

[42] and left atrium [43]. For these chambers, the assessment of diastolic function is also very important. For example, the strains for the left atrium are usually divided into reservoir strain, conduit strain, and booster strain, where the latter two occur in the diastolic phase [44]. A more accurate estimation of these diastolic strains may provide more accurate evidence for diagnosis and prognosis of various diseases."

Recently, due to the rapid advances of machine learning, many CMR-FT methods based on deep learning have been proposed [17, 45–50]. However, these learning-based feature tracking methods were almost all based on pairwise registration, and thus still suffer from the same problem as their optimization-based counterparts. On the other hand, although some learning-based groupwise registration methods have also been proposed recently for cardiac cine MRI [51–53], none of them have been applied to strain estimation. The improvement of our method over traditional pairwise feature tracking indicates that the application of learning-based groupwise registration methods could potentially yield a more accurate assessment of the strains. Additionally, learning-based strain estimation methods usually require a relatively diverse training dataset to ensure generalizability for different diseases, scanners, and image qualities. However, for some anatomies, such as the right atrium and pulmonary artery, such a diverse dataset has not been publicly available. As our method is completely unsupervised, its application to strain estimation for other chambers and vessels can be more easily available.

The study and technique have limitations. Firstly, we did not evaluate the longitudinal strain over the in vivo datasets. However, the results on the simulated dataset preliminarily validated the potential of the proposed method for longitudinal strain estimation. Secondly, we only validated our method using a relatively small patient cohort. Further evaluation based on a larger and more diverse patient cohort is highly warranted. Thirdly, we only applied our method to 2D cine images and thus could not avoid the errors caused by the out-of-plane motion.

Application of the method while exploiting the correlation between adjacent image planes may partially account for the influence of the through-plane motion [54]. Furthermore, the development of 3D cine imaging methods [55, 56] and strain estimation methods may provide an ultimate solution to this issue. Fourthly, since we did not assess the proposed method for different field strengths, slice thicknesses, flip angles, resolutions, sequences (e.g., GRE), and artifacts (e.g., severe banding and flow artifacts [57]), it is not yet known how robust the method is to images of different qualities. Although the proposed groupwise registration method may be relatively robust to low-quality images due to the exploitation of the context information of the whole image sequence, this awaits systematic investigation in large-cohort clinical studies. Finally, the computational time of the proposed method is relatively long. The typical computational time of the four CMR-FT methods for processing a single cine movie was 0.8s (Farneback optical flow), 33s (pairwise registration), 24s (Groupwise-GLR), and 38s (Groupwise-LLR). However, the computational time of the proposed method is still within a reasonable range. Further accelerations can be obtained by adopting patchwise parallel computation for the LLR dissimilarity and a more efficient optimization algorithm. Moreover, learning the complex groupwise optimization process with an end-to-end neural network may also reduce the computational time to a large extent.

## Conclusions

In conclusion, we developed a CMR-FT method based on deformable groupwise registration with an LLR dissimilarity metric. Our results based on simulated and in vivo datasets suggest that the method improves the accuracy of myocardial tracking and strain estimation compared with previous methods. Application of the proposed method may thus facilitate a more accurate assessment of the myocardial function in the clinical setting, especially along the radial direction and in the diastolic phase.

# Reference

1. Smiseth OA, Torp H, Opdahl A, Haugaa KH, Urheim S. Myocardial strain imaging: how useful is it in clinical decision making? Eur Heart J. 2016;37:1196–207.

2. Amzulescu MS, De Craene M, Langet H, Pasquet A, Vancraeynest D, Pouleur AC, et al. Myocardial strain imaging: review of general principles, validation, and sources of discrepancies. European Heart Journal - Cardiovascular Imaging. 2019;20:605–19.

3. Rajiah PS, Kalisz K, Broncano J, Goerne H, Collins JD, François CJ, et al. Myocardial Strain Evaluation with Cardiovascular MRI: Physics, Principles, and Clinical Applications. RadioGraphics. 2022;42:968–90.

4. Xu J, Yang W, Zhao S, Lu M. State-of-the-art myocardial strain by CMR feature tracking: clinical applications and future perspectives. Eur Radiol. 2022;32:5424–35.

5. Moody WE, Taylor RJ, Edwards NC, Chue CD, Umar F, Taylor TJ, et al. Comparison of magnetic resonance feature tracking for systolic and diastolic strain and strain rate calculation with spatial modulation of magnetization imaging analysis. Journal of Magnetic Resonance Imaging. 2015;41:1000–12.

6. Wu L, Germans T, Güçlü A, Heymans MW, Allaart CP, van Rossum AC. Feature tracking compared with tissue tagging measurements of segmental strain by cardiovascular magnetic resonance. Journal of Cardiovascular Magnetic Resonance. 2014;16:10.

7. Bucius P, Erley J, Tanacli R, Zieschang V, Giusca S, Korosoglou G, et al. Comparison of feature tracking, fast-SENC, and myocardial tagging for global and segmental left ventricular strain. ESC Heart Failure. 2020;7:523–32.

8. Barreiro-Pérez M, Curione D, Symons R, Claus P, Voigt J-U, Bogaert J. Left ventricular global myocardial strain assessment comparing the reproducibility of four commercially available CMR-feature tracking algorithms. Eur Radiol. 2018;28:5137–47.

9. Tobon-Gomez C, De Craene M, McLeod K, Tautz L, Shi W, Hennemuth A, et al. Benchmarking framework for myocardial tracking and deformation algorithms: An open access database. Medical Image Analysis. 2013;17:632–48.

10. Voigt J-U, Pedrizzetti G, Lysyansky P, Marwick TH, Houle H, Baumann R, et al. Definitions for a Common Standard for 2D Speckle Tracking Echocardiography: Consensus Document of the EACVI/ASE/Industry Task Force to Standardize Deformation Imaging. Journal of the American Society of Echocardiography. 2015;28:183–93.

11. De Craene M, Tobon-Gomez C, Butakoff C, Duchateau N, Piella G, Rhode KS, et al. Temporal Diffeomorphic Free Form Deformation (TDFFD) Applied to Motion and Deformation Quantification of Tagged MRI Sequences. In: Camara O, Konukoglu E, Pop M, Rhode K, Sermesant M, Young A, editors. Statistical Atlases and Computational Models of the Heart. Imaging and Modelling Challenges. Berlin, Heidelberg: Springer; 2012. p. 68–77.

12. Ye M, Kanski M, Yang D, Chang Q, Yan Z, Huang Q, et al. DeepTag: An Unsupervised Deep Learning Method for Motion Tracking on Cardiac Tagging Magnetic Resonance Images. In: 2021 IEEE/CVF Conference on Computer Vision and Pattern Recognition (CVPR). Nashville, TN, USA: IEEE; 2021. p. 7257–67.

13. Guetter C, Xue H, Chefd'hotel C, Guehring J. Efficient symmetric and inverse-consistent deformable registration through interleaved optimization. In: 2011 IEEE International Symposium on Biomedical Imaging: From Nano to Macro. Chicago, IL, USA: IEEE; 2011. p. 590–3.

14. Jolly M-P, Guetter C, Guehring J. Cardiac segmentation in MR cine data using inverse consistent deformable registration. In: 2010 IEEE International Symposium on Biomedical Imaging: From Nano to Macro. Rotterdam, Netherlands: IEEE; 2010. p. 484–7.

15. Xue H, Ding Y, Guetter C, Jolly M-P, Guehring J, Zuehlsdorff S, et al. Motion Compensated Magnetic Resonance Reconstruction Using Inverse-Consistent Deformable Registration: Application to Real-Time Cine Imaging. In: Fichtinger G, Martel A, Peters T, editors. Medical Image Computing and Computer-Assisted Intervention – MICCAI 2011. Berlin, Heidelberg: Springer Berlin Heidelberg; 2011. p. 564–72.

16. Metz CT, Klein S, Schaap M, Van Walsum T, Niessen WJ. Nonrigid registration of dynamic medical imaging data using nD+t B-splines and a groupwise optimization approach. Medical Image Analysis. 2011;15:238–49.

17. Qiao M, Wang Y, Guo Y, Huang L, Xia L, Tao Q. Temporally coherent cardiac motion tracking from cine MRI: Traditional registration method and modern CNN method. Med Phys. 2020;47:4189–98.

18. Peng Y, Ganesh A, Wright J, Xu W, Ma Y. RASL: Robust Alignment by Sparse and Low-Rank Decomposition for Linearly Correlated Images. IEEE Transactions on Pattern Analysis and Machine Intelligence. 2012;34:2233–46.

19. Haase R, Heldmann S, Lellmann J. Deformable groupwise image registration using low-rank and sparse decomposition. Journal of Mathematical Imaging and Vision. 2022;64:194–211.

20. Miao X, Lingala SG, Guo Y, Jao T, Usman M, Prieto C, et al. Accelerated cardiac cine MRI using locally low rank and finite difference constraints. Magnetic Resonance Imaging. 2016;34:707–14.

21. Vishnevskiy V, Gass T, Szekely G, Tanner C, Goksel O. Isotropic Total Variation Regularization of Displacements in Parametric Image Registration. IEEE Trans Med Imaging. 2017;36:385–95.

22. Segars WP, Sturgeon G, Mendonca S, Grimes J, Tsui BMW. 4D XCAT phantom for

multimodality imaging research. Medical Physics. 2010;37:4902–15.

23. Bernard O, Lalande A, Zotti C, Cervenansky F, Yang X, Heng P-A, et al. Deep Learning Techniques for Automatic MRI Cardiac Multi-Structures Segmentation and Diagnosis: Is the Problem Solved? IEEE Trans Med Imaging. 2018;37:2514–25.

24. Rueckert D, Sonoda LI, Hayes C, Hill DL, Leach MO, Hawkes DJ. Nonrigid registration using free-form deformations: application to breast MR images. IEEE transactions on medical imaging. 1999;18:712–21.

25. Bhatia KK, Hajnal JV, Puri BK, Edwards AD, Rueckert D. Consistent groupwise non-rigid registration for atlas construction. In: 2004 2nd IEEE International Symposium on Biomedical Imaging: Nano to Macro (IEEE Cat No. 04EX821). IEEE; 2004. p. 908–11.

26. Lai WM, Rubin D, Krempl E. Introduction to continuum mechanics. 4th ed. Amsterdam ; Boston: Butterworth-Heinemann/Elsevier; 2010.

27. Farnebäck G. Two-Frame Motion Estimation Based on Polynomial Expansion. In: Bigun J, Gustavsson T, editors. Image Analysis. Berlin, Heidelberg: Springer Berlin Heidelberg; 2003. p. 363–70.

28. Morais P, Marchi A, Bogaert JA, Dresselaers T, Heyde B, D'hooge J, et al. Cardiovascular magnetic resonance myocardial feature tracking using a non-rigid, elastic image registration algorithm: assessment of variability in a real-life clinical setting. J Cardiovasc Magn Reson. 2017;19:24.

29. Wissmann L, Santelli C, Segars WP, Kozerke S. MRXCAT: Realistic numerical phantoms for cardiovascular magnetic resonance. Journal of Cardiovascular Magnetic Resonance. 2014;16:63.

30. Selvadurai BSN, Puntmann VO, Bluemke DA, Ferrari VA, Friedrich MG, Kramer CM, et al. Definition of Left Ventricular Segments for Cardiac Magnetic Resonance Imaging. JACC: Cardiovascular Imaging. 2018;11:926–8.

31. Morais P, Heyde B, Barbosa D, Queirós S, Claus P, D'hooge J. Cardiac Motion and Deformation Estimation from Tagged MRI Sequences Using a Temporal Coherent Image Registration Framework. In: Ourselin S, Rueckert D, Smith N, editors. Functional Imaging and Modeling of the Heart. Berlin, Heidelberg: Springer Berlin Heidelberg; 2013. p. 316–24.

32. Cohen J. Statistical Power Analysis for the Behavioral Sciences. 2nd edition. New York: Routledge; 1988.

33. Koo TK, Li MY. A Guideline of Selecting and Reporting Intraclass Correlation Coefficients for Reliability Research. J Chiropr Med. 2016;15:155–63.

34. Luetkens JA, Schlesinger-Irsch U, Kuetting DL, Dabir D, Homsi R, Doerner J, et al. Feature-tracking myocardial strain analysis in acute myocarditis: diagnostic value and association with myocardial oedema. Eur Radiol. 2017;27:4661–71.

35. Fischer K, Obrist SJ, Erne SA, Stark AW, Marggraf M, Kaneko K, et al. Feature Tracking Myocardial Strain Incrementally Improves Prognostication in Myocarditis Beyond Traditional CMR Imaging Features. JACC: Cardiovascular Imaging. 2020;13:1891–901.

36. Militaru S, Panovsky R, Hanet V, Amzulescu MS, Langet H, Pisciotti MM, et al. Multivendor comparison of global and regional 2D cardiovascular magnetic resonance feature tracking strains vs tissue tagging at 3T. Journal of Cardiovascular Magnetic Resonance. 2021;23:54.

37. Pandey T, Alapati S, Wadhwa V, Edupuganti MM, Gurram P, Lensing S, et al. Evaluation of Myocardial Strain in Patients With Amyloidosis Using Cardiac Magnetic Resonance Feature Tracking. Current Problems in Diagnostic Radiology. 2017;46:288–94.

38. Wisotzkey BL, Soriano BD, Albers EL, Ferguson M, Buddhe S. Diagnostic role of strain imaging in atypical myocarditis by echocardiography and cardiac MRI. Pediatr Radiol. 2018;48:835–42.

39. Kim MY, Park E-A, Lee W, Lee S-P. Cardiac Magnetic Resonance Feature Tracking in Aortic Stenosis: Exploration of Strain Parameters and Prognostic Value in Asymptomatic Patients with Preserved Ejection Fraction. Korean J Radiol. 2020;21:268.

40. Seemann F, Baldassarre LA, Llanos-Chea F, Gonzales RA, Grunseich K, Hu C, et al. Assessment of diastolic function and atrial remodeling by MRI – validation and correlation with echocardiography and filling pressure. Physiological Reports. 2018;6:e13828.

41. He J, Yang W, Wu W, Li S, Yin G, Zhuang B, et al. Early Diastolic Longitudinal Strain Rate at MRI and Outcomes in Heart Failure with Preserved Ejection Fraction. Radiology. 2021;301:582–92.

42. Erley J, Tanacli R, Genovese D, Tapaskar N, Rashedi N, Bucius P, et al. Myocardial strain analysis of the right ventricle: comparison of different cardiovascular magnetic resonance and echocardiographic techniques. Journal of Cardiovascular Magnetic Resonance. 2020;22:51.

43. Cau R, Bassareo P, Suri JS, Pontone G, Saba L. The emerging role of atrial strain assessed by cardiac MRI in different cardiovascular settings: an up-to-date review. Eur Radiol. 2022;32:4384–94.

44. Song Y, Li L, Chen X, Shao X, Lu M, Cheng J, et al. Early Left Ventricular Diastolic Dysfunction and Abnormal Left Ventricular-left Atrial Coupling in Asymptomatic Patients With Hypertension: A Cardiovascular Magnetic Resonance Feature Tracking Study. Journal of Thoracic Imaging. 2022;37:26.

45. Morales MA, Van den Boomen M, Nguyen C, Kalpathy-Cramer J, Rosen BR, Stultz CM, et al. DeepStrain: a deep learning workflow for the automated characterization of cardiac mechanics. Frontiers in Cardiovascular Medicine. 2021;:1041.

46. Qin C, Wang S, Chen C, Bai W, Rueckert D. Generative myocardial motion tracking via latent space exploration with biomechanics-informed prior. Medical Image Analysis. 2023;83:102682.

47. V. Graves C, Rebelo MFS, Moreno RA, Dantas-Jr RN, Assunção-Jr AN, Nomura CH, et al. Siamese pyramidal deep learning network for strain estimation in 3D cardiac cine-MR. Computerized Medical Imaging and Graphics. 2023;108:102283.

48. Yu H, Chen X, Shi H, Chen T, Huang TS, Sun S. Motion Pyramid Networks for Accurate and Efficient Cardiac Motion Estimation. In: Martel AL, Abolmaesumi P, Stoyanov D, Mateus D, Zuluaga MA, Zhou SK, et al., editors. Medical Image Computing and Computer Assisted Intervention – MICCAI 2020. Cham: Springer International Publishing; 2020. p. 436–46.

49. Yu H, Sun S, Yu H, Chen X, Shi H, Huang TS, et al. FOAL: Fast Online Adaptive Learning for Cardiac Motion Estimation. In: 2020 IEEE/CVF Conference on Computer Vision and Pattern Recognition (CVPR). Seattle, WA, USA: IEEE; 2020. p. 4312–22.

50. Pan J, Rueckert D, Küstner T, Hammernik K. Efficient Image Registration Network for Non-Rigid Cardiac Motion Estimation. In: Haq N, Johnson P, Maier A, Würfl T, Yoo J, editors. Machine Learning for Medical Image Reconstruction. Cham: Springer International Publishing; 2021. p. 14–24.

51. Martín-González E, Sevilla T, Revilla-Orodea A, Casaseca-de-la-Higuera P, Alberola-López C. Groupwise Non-Rigid Registration with Deep Learning: An Affordable Solution Applied to 2D Cardiac Cine MRI Reconstruction. Entropy. 2020;22:687.

52. Hammernik K, Pan J, Rueckert D, Küstner T. Motion-Guided Physics-Based Learning for Cardiac MRI Reconstruction. In: 2021 55th Asilomar Conference on Signals, Systems, and Computers. 2021. p. 900–7.

53. Pan J, Hamdi M, Huang W, Hammernik K, Kuestner T, Rueckert D. Unrolled and rapid motion-compensated reconstruction for cardiac CINE MRI. Medical Image Analysis. 2024;91:103017.

54. Hess AT, Zhong X, Spottiswoode BS, Epstein FH, Meintjes EM. Myocardial 3D strain calculation by combining cine displacement encoding with stimulated echoes (DENSE) and cine strain encoding (SENC) imaging. Magnetic Resonance in Med. 2009;62:77–84.

55. Usman M, Ruijsink B, Nazir MS, Cruz G, Prieto C. Free breathing whole-heart 3D CINE MRI with self-gated Cartesian trajectory. Magnetic Resonance Imaging. 2017;38:129–37.

56. Küstner T, Fuin N, Hammernik K, Bustin A, Qi H, Hajhosseiny R, et al. CINENet: deep learning-based 3D cardiac CINE MRI reconstruction with multi-coil complex-valued 4D spatio-temporal convolutions. Sci Rep. 2020;10:13710.

57. Chen Z, Hua S, Gao J, Chen Y, Gong Y, Shen Y, et al. A dual-stage partially interpretable neural network for joint suppression of bSSFP banding and flow artifacts in non-phase-cycled cine imaging. Journal of Cardiovascular Magnetic Resonance. 2023;25:68.